# CoFeB Thickness Dependence of Thermal Stability Factor in CoFeB/MgO Perpendicular Magnetic Tunnel Junctions

H. Sato[1,a)], M. Yamanouchi[1], K. Miura[1,2,3], S. Ikeda[1,2], R. Koizumi[2], F. Matsukura[1,2] and H. Ohno[1,2]

[1]*Center for Spintronics Integrated Systems, Tohoku University, 2-1-1 Katahira, Aoba-ku, Sendai, 980-8577, Japan*

[2]*Laboratory for Nanoelectronics and Spintronics, Research Institute of Electrical Communication, Tohoku University, 2-1-1 Katahira, Aoba-ku, Sendai 980-8577, Japan*

[3]*Central Research Laboratory, Hitachi, Ltd., Kokubunji, Tokyo 185-8601, Japan*

Abstract

Thermal stability factor $\Delta$ of recording layer was studied in perpendicular anisotropy CoFeB/MgO magnetic tunnel junctions (p-MTJs) with various CoFeB recording layer thicknesses and junction sizes. In all series of p-MTJs with different thicknesses, $\Delta$ is virtually independent of the junction sizes of 48–81 nm in diameter. The values of $\Delta$ increase linearly with increasing the recording layer thickness. The slope of the linear fit is explained well by a model based on nucleation type magnetization reversal.

Magnetic tunnel junctions (MTJs) with spin transfer torque (STT) switching have attracted much attention for use in nonvolatile random access memory[1-4] and spintronics based logic-in-memory.[5] The MTJs need to satisfy the following characteristics simultaneously to be used with complementary metal-oxide-semiconductor (CMOS) integrated circuits; a high tunnel magnetoresistance (TMR) ratio over 100%, a switching current lower than the corresponding transistor drive current, a high thermal stability for sufficient retention time, and annealing treatment stability at 350–400°C for back end process. MgO based MTJs have high potential for integration with CMOS circuits because of their high TMR ratio, predicted theoretically[6,7] and subsequently demonstrated by several groups.[8-11] Additionally, CoFeB/MgO MTJs with perpendicular anisotropy (p-MTJs), which is brought forth by the interface anisotropy at CoFeB-MgO,[12] have been shown to have a high potential to meet these requirements simultaneously.[13] Moreover, it was reported that CoFeB/MgO p-MTJs exhibit more preferable characteristics (faster STT switching speed[14] and lower STT write error rate[15]) for integrated circuit applications than CoFeB/MgO MTJs with in-plane anisotropy.

In CoFeB/MgO p-MTJs, intrinsic critical current ($I_{C0}$) was found to increase proportionally with increasing junction area, whereas the thermal stability factor ($\Delta = E/k_BT$, where $E$ denotes the energy barrier between the two magnetization configurations of MTJ, $k_B$ the Boltzmann constant, and $T$ the absolute temperature = 300K) was almost constant between the junction size of 40–78 nm in diameter.[16] Such behavior was attributed to nucleation type magnetization reversal, as discussed similarly in perpendicular patterned media.[17-22] In perpendicular patterned media, it was shown experimentally that the nucleation size in diameter is comparable to the domain wall width ($\delta_w = \pi(A_s/K_{eff})^{1/2}$, where $A_s$ is the exchange stiffness constant and $K_{eff}$ is the effective magnetic anisotropy energy).[22] Moreover, the estimated nucleation size in our CoFeB/MgO p-MTJs was close to the $\delta_w$ determined in a separate study.[23] These experimentally obtained results imply that $E$ in $\Delta$ and the magnetic recording layer thickness ($t_{rec}$) can be related to each other for MTJs having diameter larger than $\delta_w$ as

$$E \approx K_{eff}\pi(\delta_w/2)^2 t_{rec} = \pi^3 A_s t_{rec}/4 . \quad (1)$$

If this is indeed the case, then $\Delta$ is expected to increase proportionally with $t_{rec}$, and it is independent of $K_{eff}$. Sun *et al*. also pointed out that $\Delta$ is proportional to $A_s t_{rec}$, considering magnon excitations.[24] Here, we studied $\Delta$ of CoFeB/MgO p-MTJs with various CoFeB recording layer thicknesses and junction sizes to address the relationship between $\Delta$ and $t_{rec}$ experimentally.

Stacking structures consisting of, from the substrate side, Ta(5)/Ru(10)/Ta(5)/Co$_{20}$Fe$_{60}$B$_{20}$(0.9)/MgO(0.9)/Co$_{20}$Fe$_{60}$B$_{20}$($t_{CoFeB}$)/Ta(5)/Ru(5) (numbers in parenthesis are nominal thicknesses in nm) were deposited on thermally oxidized Si wafer using rf magnetron sputtering. Nominal CoFeB layer thickness ($t_{CoFeB}$) was 1.3, 1.5, 1.6, or 1.7 nm. MTJs were fabricated using electron beam lithography and Ar ion milling followed by annealing at 300°C for 1 hour in vacuum under a magnetic field of 0.4 T along the direction perpendicular to the film plane. Circular CoFeB/MgO p-MTJs of 48–81 nm in diameter were fabricated. Junction sizes of the MTJs were defined as the physical area determined from scanning electron microscope (SEM) images of MTJs. Figure 1 shows the TMR ratio for the CoFeB/MgO p-MTJs as a function of $t_{CoFeB}$. The TMR ratio was virtually the same within the $t_{CoFeB}$ range studied here. Similar results were obtained for all series of MTJs studied. The reference or recording layer assignment was done based on the current direction of STT switching with respect to the magnetization configuration (parallel (P) or antiparallel (AP) states); the bottom (top) CoFeB layer was determined as the reference (recording) layer.

To obtain $\Delta$ of the recording layer for each MTJ, we measured the switching probability $P(\tau)$ of the recording layer as a function of the magnetic field amplitude and evaluated $\Delta$ using the following relationship based on the Stoner–Wohlfarth model,[25] as shown below.

$$P(\tau) = 1 - \exp\left[-\frac{\tau}{\tau_0}\exp\left\{-\Delta(1-\frac{H-H_s}{H_k^{eff}})^2\right\}\right] \qquad (2)$$

In that equation, $\tau$ denotes the pulse magnetic field duration (1 s in this study), $\tau_0$ the inverse of attempt frequency (assumed to be 1 ns), $H_k^{eff}$ the effective magnetic anisotropy field, and $H$ the external magnetic field amplitude, which is equal to the sum of the pulse magnetic field $H_p$ and the dc magnetic field $H_{DC}$. $H_S$ (>0 under the experimental condition employed here) is the shift field of the resistance-field curves, *i.e.* the field between the center of the recording minor $R$–$H$ curve and zero magnetic field (dipole magnetic fields from the reference layer are the major source[13]). A typical switching probability for a $t_{CoFeB} = 1.6$ nm CoFeB/MgO p-MTJ of 59 nm in diameter is presented in Fig. 2. In Fig. 2(a), minor $R$–$H$ curves measured 100 times are shown, from which the switching probability is obtained as a function of the external magnetic field ($\mu_0 H = \mu_0 H_p + \mu_0 H_{DC}$), as displayed in Fig. 2(b). Symbols presented in Fig. 2(b) correspond to the experimentally obtained results and the solid line is a fit calculated from Eq. (2), from which $H_s$, $H_k^{eff}$, and $\Delta$ are obtained.

Figure 3 shows $\Delta$ for the MTJs with $t_{CoFeB} = 1.3-1.7$ nm as a function of the recording layer area. The measured $\Delta$'s are insensitive to the recording layer area in all series of MTJs studied here, which indicates that such $\Delta$ behavior with respect to the recording layer area is a general phenomenon for the CoFeB thicknesses studied here, *i.e.* within the range of thickness that results in perpendicular magnetic Dotted lines in these figures correspond to the average $\Delta$ for each series. Parenthetically, $\Delta$ of MTJs with $t_{CoFeB}$ of 1.3 nm and diameter of 48 nm$\phi$ could not be measured because of its thermal instability of the AP state. Figure 4 shows $H_s$ for CoFeB/MgO p-MTJs with $t_{CoFeB} = 1.3-1.7$ nm as a function of the recording layer area. $H_s$ increases continuously with decreasing the recording layer area, which is consistent with previously reported results.[26]

We now focus on the relationship between the average $\Delta$'s obtained in Fig. 3 and the actual CoFeB recording layer thickness ($t_{CoFeB}^*$). $t_{CoFeB}^*$ ($= t_{rec}$) is obtained by subtracting magnetically dead layer thickness at the CoFeB/Ta interface from the nominal thickness $t_{CoFeB}$.[13] The magnetically dead layer (0.4 nm $\pm$ 0.1 nm under the present condition) was determined from the thickness dependence of magnetic moment per unit area in separately deposited layers consisting of Ta(5)/MgO(0.9)/Co$_{20}$Fe$_{60}$B$_{20}$($t_{CoFeB} = 0.9$-5)/Ta(5)/Ru(5) annealed under the same condition. As can be seen in Fig. 5, a linear relationship exists between $\Delta$ and $t_{CoFeB}*$. This is in accordance with Eq. (1), which was inferred from the experimentally obtained results reported previously.[16, 22] Although Eq. (1) is approximate, the linear relationship observed in the experiment is sufficiently tempting to extract an effective exchange stiffness constant $A_s*$ for the $t_{CoFeB}*$ range of 0.9–1.3 nm. From the slope of 36 nm$^{-1}$, one obtains $A_s* \approx 19$ pJ/m, a value close to 31 pJ/m obtained from the domain periodicity measurements of a Ta(5)/Ru(10)/Ta(5)/CoFeB(1.3)/MgO(1)/Ta(2) stack structure annealed at 350°C.[23] It is noteworthy that the effective exchange stiffness constant obtained from the slope of linear fit in Fig. 5 is very close to the exchange stiffness constant values for CoFeB films reported previously.[27,28] To confirm the consistency, we also calculate the length scale responsible for the nucleation $\delta_w*$ using $A_s*$ and $K_{eff}$ obtained from $K_{eff} = H_k^{eff}M_s/2$, where $H_k^{eff}$ is determined from the switching probability measurements and $M_s$ from the vibrating sample magnetometer for continuous CoFeB films. The $\delta_w*$ values are equal to 60, 40, 39, and 45 nm for $t_{CoFeB}* = 0.9, 1.1, 1.2, 1.3$ nm, respectively, which are comparable to the smallest junction sizes at each $t_{CoFeB}*$ studied here.

Finally, the influence of dipolar coupling on $\Delta$ is discussed. As shown in Eq. (2), an increase of $H_s$ reduces thermal stability at the AP state, which was verified experimentally in CoFeB/MgO p-MTJs annealed at 400°C.[29] The absolute value of $H_s$ is determined experimentally as 30–40 mT at a dimension of 48 nm$\phi$. However, this value increases continuously with decreasing recording layer area, as shown in Fig. 4, which causes thermal instability of the AP state in a smaller dimension. Moreover, $H_s$ affects the switching ability by STT. CoFeB/MgO p-MTJs with $t_{CoFeB}$ of more than 1.5 nm show clear switching events by STT. However, CoFeB p-MTJs with $t_{CoFeB} = 1.3$ nm show switching both from AP to P and P to AP state in the positive current region, because of AP state instability in a zero magnetic field resulting from dipole coupling.

In summary, we investigated the CoFeB recording layer thickness dependence of $\Delta$ to experimentally explore the relationship between $\Delta$ and $t_{CoFeB}$*. Results show that $\Delta$ with junction size of 48–81 nm in diameter is insensitive to the junction area. On the other hand, $\Delta$'s increase linearly with increasing $t_{CoFeB}$*, independent of the anisotropy $K_{eff}$, which can be explained by the nucleation type magnetization reversal scenario, indicating that the $\Delta$ of CoFeB/MgO perpendicular MTJs is governed by the exchange stiffness $A_s$* and $t_{CoFeB}$*. Although the relationship found in this study is a step forward to full understanding of $\Delta$, we point out that in a structure with reduced dipole fields, we observed $\Delta$ of 70,[30] which is larger than the highest value observed in this study. This observation indicates that other factors influence $\Delta$. Further studies are necessary to fully establish an understanding of $\Delta$.

The work was supported by the FIRST program of JSPS and the GCOE program at Tohoku University. The authors wish to thank R. Sasaki, I. Morita, T. Hirata, H. Iwanuma for their technical support in MTJ fabrication.

**References**

[1]M. Hosomi, H. Yamagishi, T. Yamamoto, K. Bessho, Y. Higo, K. Yamane, H. Yamada, M. Shoji, H. Hachino, C. Fukumoto, H. Nagao, and H. Kano, Tech. Dig. Int. Electron Devices Meet. 2005, p.459.

[2]T. Kawahara, R. Takemura, K. Miura, J. Hayakawa, S. Ikeda, Y. M. Lee, R. Sasaki, Y. Goto, K. Ito, T. Meguro, F. Matsukura, H. Takahashi, H. Matsuoka, and H. Ohno, IEEE J. Solid-State Circuits **43**, 109 (2008).

[3]R. Takemura, T. Kawahara, K. Miura, H. Yamamoto, J. Hayakawa, N. Matsuzaki, K. Ono, M. Yamanouchi, K. Ito, H. Takahashi, S. Ikeda, H. Hasegawa, H. Matsuoka, and H. Ohno, IEEE J. Solid-State Circuits **45**, 869 (2010).

[4]T. Kishi, H. Yoda, T. Kai, T. Nagase, E. Kitagawa, M. Yoshikawa, K. Nishiyama, T. Daibou, M. Nagamine, M. Amano, S. Takahashi, M. Nakayama, N. Shimomura, H. Aikawa, S. Ikegawa, S. Yuasa, K. Yakushiji, H. Kubota, A. Fukushima, M. Oogane, T. Miyazaki, and K. Ando, IEEE International Electron Devices Meeting (IEDM) Technical Digest 2008, p. 309.

[5]S. Matsunaga, J. Hayakawa, S. Ikeda, K. Miura, H. Hasegawa, T. Endoh, H. Ohno, and T. Hanyu: Appl. Phys. Express **1**, 091301 (2008).

[6]W. H. Butler, X. G. Zhang, T. C. Schlthess, J. M. MacLaren, Phys. Rev. B., **63**, 054416 (2001).

[7]J. Mathon and A. Umerski, Phys. Rev., B., **63**, 220403(R) (2001).

[8]S. Yuasa, T. Nagahama, A. Fukushima, Y. Suzuki, and K. Ando, Nature. Mater., **3**, 868 (2004).

[9]S. S. P. Parkin, C. Kaiser, A. PanchulaA, P. M. Rice, B. Hughes, M. Samant, and S. H. Yang, Nature. Mater., **3**, 862 (2004).

[10]D. D. Djayaprawira, K. Tsunekawa, M. Nagai, H. Maehara, S. Yamagata, N. Watanabe, S. Yuasa, Y. Suzuki, and K. Ando, Appl. Phys. Lett., **86**, 092502 (2005).

[11]J. Hayakawa, S. Ikeda, F. Matsukura, H. Takahashi, and H. Ohno, Jpn. J. Appl. Phys., Part 2, **44**, L587 (2005).

[12]M. Endo, S. Kanai, S. Ikeda, F. Matsukura and H. Ohno, Appl. Phys. Lett. **96**, 212503 (2010).

[13]S. Ikeda, K. Miura, H. Yamamoto, K. Mizunuma, H. D. Gan, M. Endo, S. Kanai, J. Hayakawa, F. Matsukura and H. Ohno, Nat. Mat. **9**, 721 (2010).

[14]D. C. Worledge, G. Hu, David W. Abraham, J. Z. Sun, P. L. Trouilloud, J. Nowak, S. Brown, M. C. Gaidis, J. O'Sullivan and R P. Robertazzi, Appl. Phy. Lett. **98**, 022501 (2011).

[15]J. J. Nowak, R. P. Robertazzi, J. Z. Sun, G. Hu, D. W. Abraham, P. L. Trouilloud, S. Brown, M. C. Gaidis, E. J. O'Sullizan, W. J. Gallagher and D. C. Worledge, IEEE Magn. Lett. **2**, 3000204 (2011).

[16]H. Sato, M. Yamanouchi, K. Miura, S. Ikeda, H. D. Gan, K. Mizunuma, R. Koizumi, F. Matsukura and H. Ohno, Appl. Phys. Lett. **99**, 042501 (2011).

[17]C. T. Rettner S. Anders, T. Thomson, M. Albrecht, Y. Ikeda, M. E. Best, and B. D. Terris, IEEE. Trans. Magn., **38**, 1725 (2002).

[18]S. P. Li, A. Lelib, Y. Chen, Y. Fu and M. E. Welland, J. Appl. Phys. **91**, 9964 (2002).

[19]J. Moritz, B. Dieny, J. P. Nozieres, R. J. M. Veerdonk, T. M. Crawford, D. Weller, and S. Landis, Appl. Phys. Lett., **86**, 063512 (2005).

[20]K. Mitsuzuka, N. Kikuchi, T. Shimatsu, O. Kitakami, H. Aoi, H. Muraoka, and J. C. Lodder, IEEE. Trans. Magn., **43**, 2160 (2007).

[21]S. Okamoto, T. Kato, N. Kikuchi, O. Kitakami, N. Tezuka, and S. Sugimoto, J. Appl. Phys., **103**, 07C501 (2008).

[22]N. Kikuchi, K. Mitsuzuka, T. Shimatsu, O. Kitakami, and H. Aoi, J. Phys. D, **200**, 102003

(2010).

[23]M. Yamanouchi, A. Jander, P. Dhagat, S. Ikeda, F. Matsukura and H. Ohno, IEEE Magn. Lett. **2**, 3000304 (2011).

[24]J. Z. Sun, R. P. Robertazzi, J. Nowak, P. L. Trouilloud, G. Hu, D. W. Abraham, M. C. Gaidis, S. L. Brown, E. J. O'Sullivan, W. J. Gallagher and D. C. Worledge, Phy. Rev. B **84**, 064413 (2011).

[25]Z. Li and S. Zhang, Phys. Rev. B. **69**, 134416 (2004).

[26]S. Bandiera, R. C. Sousa, Y. Dahmane, D. Ducruet, C. Portemont, V. Baltz, S. Auffret, I. L. Prejbeanu, and B. Dieny, IEEE Magn. Lett., **1**, 3000204 (2010).

[27]C. Bilzer, T. Devolder, J. V. Kim, G. Counil, C. Chappert, C. Cardoso, and P. P. Freitas, J, Appl. Phys., **100**, 053903 (2006).

[28]A. Helmer, S. Cornelissen, T. Devolder, J. V. Kim, W. V. Roy, L. Lagae, and C. Chappert, Phys. Rev. B, **81**, 094416 (2010).

[29]H. D. Gan, H. Sato, S. Ikeda, M. Yamanouchi, K. Miura, R. Koizumi, F. Matsukura, and H. Ohno, Appl. Phys. Lett., **99**, 252507 (2011).

[30]K. Miura, S. Ikeda, M. Yamanouchi, H. Yamamoto, K. Mizunuma, H. D. Gan, J. Hayakawa, R. Koizumi, F. Matsukura, and H. Ohno, Dig. Tech. Pap. – Symp. VLSI Technol. **2011**, 214.

**Figure captions**

FIG. 1. Average TMR ratio of MTJs with 48-81 nm$\phi$ is shown as a function of nominal recording layer thickness $t_{CoFeB}$.

FIG. 2. Typical switching probability measurement result for $t_{CoFeB}$ = 1.6 nm MTJ of 59 nm in diameter. (a) Resistance vs. applied magnetic field (pulse magnetic field duration = 1 s) curves measured under 1 μA sense current for resistance measurement. (b) External magnetic field dependence of switching probability for the P to AP (AP to P) state obtained from the result shown in (a).

FIG. 3. $\varDelta$ for $t_{CoFeB}$ = 1.3 nm (a), 1.5 nm (b), 1.6 nm (c), 1.7 nm (d) MTJs as a function of the recording layer area. Smallest and largest junction sizes estimated from SEM image for each series of MTJs are also shown in these figures. Dotted lines in these figures correspond to average $\varDelta$ for all MTJs in each series of CoFeB/MgO p-MTJs with different $t_{CoFeB}$.

Fig. 4 $H_s$ for CoFeB/MgO p-MTJs with $t_{CoFeB}$ of 1.3−1.7 nm as a function of the recording layer area.

FIG. 5. Averaged $\varDelta$'s for MTJs of 48−81 nm diameter as a function of $t_{CoFeB}$*.

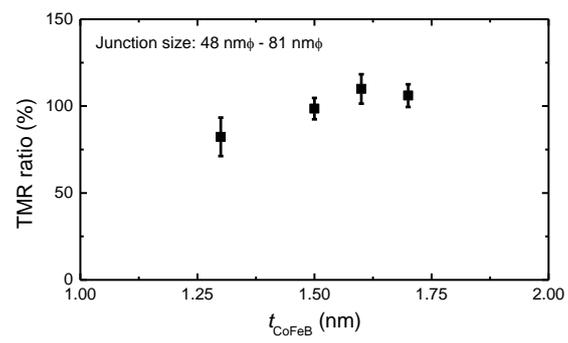

Figure 1

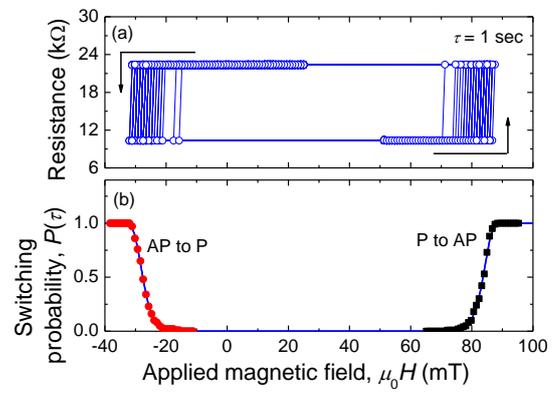

Figure 2

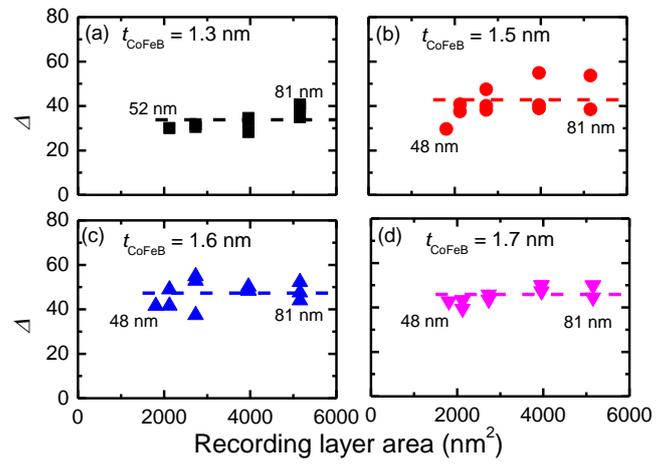

Figure 3

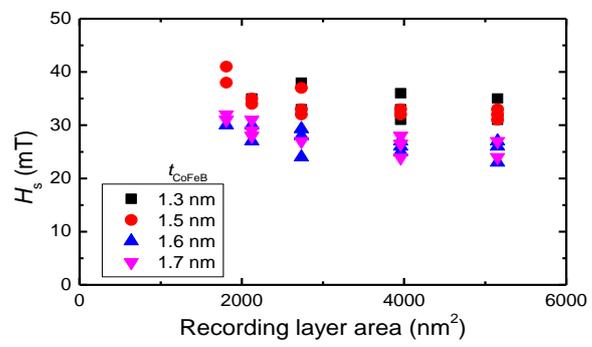

Figure 4

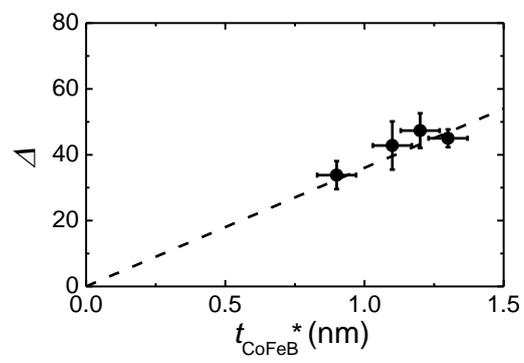

Figure 5